\documentclass[11pt,a4paper]{article}
\usepackage{jheppub}

\usepackage{graphicx}
\usepackage{amsmath}
\usepackage{epsfig,multicol,bbm}

\newcommand{\be}{\begin{equation}}
\newcommand{\ee}{\end{equation}}
\newcommand{\bea}{\begin{eqnarray}}
\newcommand{\eea}{\end{eqnarray}}

\newcommand{\mc}{\mathcal}

\newcommand{\lsim}{\lesssim}

\newcommand{\beqa}{\begin{eqnarray}}
\newcommand{\eeqa}{\end{eqnarray}}

\newcommand{\vo}{\mathcal{V}}

\newcommand\fverb{\setbox\fverbbox=\hbox\bgroup\verb}
\newcommand\fverbdo{\egroup\medskip\noindent
			\fbox{\unhbox\fverbbox}\ }
\newcommand\fverbit{\egroup\item[\fbox{\unhbox\fverbbox}]}
\newbox\fverbbox

\begin{document}

\title{Extended No-Scale Structure and $\alpha^{'2}$ Corrections to the Type IIB Action}

\abstract{We analyse a new ${\cal N}=1$ string tree level correction at ${\cal O}(\alpha'^2)$ to the K\"ahler potential of the volume moduli of type IIB Calabi-Yau flux compactification found recently by Grimm, Savelli and Weissenbacher~\cite{Grimm:2013gma} and its impact on the moduli potential. We find that it imposes a strong lower bound the Calabi-Yau volume in the Large Volume Scenario of moduli stabilisation. For KKLT-like scenarios we find that consistency of the action imposes an upper  bound  on the flux superpotential $|W_0|\lesssim 10^{-3}$, while parametrically controlled survival of the KKLT minimum needs extreme tuning of $W_0$ close to zero. We also analyse the K\"ahler uplifting mechanism showing that it can operate on Calabi-Yau manifolds where the new correction is present and dominated by the 4-cycle controlling the overall volume if the volume is stabilised at values $\vo \gtrsim 10^3$. We discuss the phenomenological implication of these bounds on $\vo$ in the various scenarios.
}

\preprint{DESY-13-087, NSF-KITP-13-085}

\author[a]{F. G. Pedro,}
\author[b,c]{M. Rummel,}
\author[a,d]{A. Westphal}

\affiliation[a]{Deutsches Elektronen-Synchrotron DESY, Theory Group, D-22603 Hamburg, Germany}
\affiliation[b]{II. Institut f\"ur Theoretische Physik der Universit\"at Hamburg, D-22761 Hamburg, Germany}
\affiliation[c]{Institute for Advanced Study, Hong Kong University of Science and Technology, Hong Kong}
\affiliation[d]{Kavli Institute for Theoretical Physics, Santa Barbara, California 93106, USA}
\emailAdd{francisco.pedro@desy.de}\emailAdd{markus.rummel@desy.de}\emailAdd{alexander.westphal@desy.de}

\maketitle

\section{Introduction}

String theory is a candidate fundamental theory of quantum gravity in unification with the three non-gravitational forces of the Standard Model (SM) of particle physics. If we interpret its 2D worldsheet conformal field theory in terms of geometric embedding space-time, then string theory requires ten-dimensional backgrounds. Hence, contact with the 4D nature of space-time at low energies requires compactification of the six extra dimensions on a compact manifold, typically with a high-dimensional deformation or 'moduli' space. The requirement of 4D ${\cal N}=1$ supersymmetry then selects a specific class of 6-manifolds of which Calabi-Yau 3-folds (CY3s) comprise a large and rich subset.

Moreover, CY compactifications of string theory must describe both broken supersymmetry below a TeV, and the presence of an extremely small but positive vacuum energy driving late-time accelerated cosmic expansion, if they are to make contact with 4D low energy physics. This requires moduli stabilisation in combination with spontaneous breaking of 4D ${\cal N}=1$ supersymmetry (SUSY) in a local near-Minkowski minimum of the moduli scalar potential.

Recent years have seen considerable progress in this direction, using combinations of higher p-form gauge fluxes of string theory with orientifold planes, non-perturbative effects from D-branes, and perturbative corrections to the kinetic terms of the moduli scalar fields.\footnote{For reviews of flux compactifications see e.g., \cite{Douglas:2006es,Grana:2005jc,Blumenhagen:2006ci}.} Various setups involving these effects can lead to a large discretum of isolated meta-stable anti-de Sitter (AdS), Minkowski, or de Sitter (dS) minima of the moduli potential on a given CY compactification, giving rise to the so-called string theory landscape~\cite{Kachru:2003aw,Susskind:2003kw}.

In particular, type II string theories are amenable to this array of techniques as they comprise a rich set of both RR fluxes in addition to the universal NSNS $H_3$ 3-form flux. In the context of type IIB string theory a combination of NSNS and RR 3-form stabilises supersymmetrically all complex structure moduli of the CY3 as well as the axio-dilaton containing the string coupling~\cite{Giddings:2001yu,Dasgupta:1999ss}. There are several avenues available to stabilise the K\"ahler or volume moduli of the CY 3-fold. Firstly, one can use a combination of perturbative corrections to the K\"ahler potential of the K\"ahler moduli to generate non-SUSY AdS minima~\cite{Saltman:2004jh,Parameswaran:2006jh,Silverstein:2007ac}. Alternatively, one can use a combination of perturbative K\"ahler corrections, and of non-perturbative effects in the superpotential from gaugino condensation of non-Abelian gauge theories on stacks of D7-branes, or from Euclidean D3-branes, wrapping 4-dimensional subspaces of the CY3 (the 4-
cycles) to stabilise the K\"ahler moduli measuring the 4-cycle volumes. Here there are three possibilities in the extant literature:
\begin{itemize}
\item If we use solely non-perturbative effects in the superpotential, we end up with SUSY AdS vacua for the K\"ahler moduli~\cite{Kachru:2003aw}. Consequently, we need to introduce an additional source of SUSY breaking and 'uplifting' to near-zero vacuum energy. This may be an anti-D3-brane at the tip of a warped throat region of the CY~\cite{Kachru:2003aw}, or a combination of F- and D-terms generated by additional 'matter' field sectors~\cite{Burgess:2003ic,Lebedev:2006qq}.
\item If the CY has at least 2 K\"ahler moduli, a combination of the non-perturbative effects from Euclidean D3-branes or gaugino condensation on D7-branes and the perturbative ${\cal O}(\alpha'^3)$-correction to the volume moduli K\"ahler potential can produce non-SUSY AdS vacua at exponentially large CY volumes in the Large Volume Scenario (LVS)~\cite{Balasubramanian:2005zx}. The ${\cal O}(\alpha'^3)$-correction to the volume moduli K\"ahler potential arises from a 10D $R^4$ string theory correction to the 10D effective type IIB supergravity action~\cite{Becker:2002nn}. An additional source of vacuum energy is necessary to reach zero vacuum energy.
\item In 'K\"ahler uplifting', we can play off the ${\cal O}(\alpha'^3)$-correction to the volume moduli K\"ahler potential and a non-perturbative superpotential from gaugino condensation on stacked D7-branes in a different way than in the LVS~\cite{Balasubramanian:2004uy,Rummel:2011cd,Louis:2012nb}. In this setup we can achieve direct stabilisation of one or more K\"ahler moduli into a minimum at moderately large volume with adjustable vacuum energy, both AdS or dS.
\end{itemize}

Given these existing paths to moduli stabilisation, it is critically important to check for the existence of additional, potentially leading,  perturbative corrections to the volume moduli K\"ahler potential of type IIB CY compactifications. While such corrections can arise at ${\cal O}(\alpha'^2g_s^2)$ from string loop corrections~\cite{Berg:2005ja,Berg:2007wt,Cicoli:2007xp}, they contribute in the resulting scalar potential only at ${\cal O}(\alpha'^4)$ due to the extended no-scale structure of the type IIB supergravity moduli potential~\cite{Berg:2007wt,Cicoli:2007xp}. This feature has its origin in the structure of the string loop corrections which appear as homogeneous degree-minus-2 polynomials in the 2-cycle volumes in the K\"ahler potential of the volume moduli. As the string loop corrections are $g_s$-suppressed and appear generically with ${\cal O}(1)$ coefficients, they preserve the three previous vacuum constructions.

However, recently Grimm, Savelli and Weissenbacher (GSW)~\cite{Grimm:2013gma} derived for the first time the presence of a manifestly ${\cal N}=1$ string tree level ${\cal O}(\alpha'^2)$ correction from a higher derivative M-theory correction which is transported to F-theory and finally to its weak coupling limit, i.e., type IIB string theory. This correction is controlled by the volume of the intersection between the D7-branes and the O7-plane, which is a linear combination of 2-cycle volumes. It is not suppressed by powers of $g_s$ and typically appears with a prefactor larger than ${\cal O}(1)$. Despite sharing the extended no-scale structure with the string loop corrections, the large prefactors combined with a just marginally stronger suppression by a factor $\vo^{-1/3}$ in the scalar potential may spell danger to some or all of the above mechanism of moduli stabilisation.  Consequently, we analyse the effect of the GSW correction on all three models of non-perturbative, or mixed perturbative-non-
perturbative K\"ahler moduli stabilisation. 

The outline of the paper is as follows. In Section~\ref{sec:ex-no-scale} we review the extended no-scale structure of the supergravity scalar potential for the K\"ahler moduli of type IIB CY flux compactifications, and determine the leading terms induced by the GSW correction. In Section~\ref{sec:GS} we discuss the structure of the GSW correction in the 4D volume moduli K\"ahler potential for general CY 3-folds with several K\"ahler moduli, in particular for CYs with a classical volume of approximate `swiss-cheese' form. Section~\ref{sec:KKLT} analyses the scalar potential for a general 1-parameter manifold both in the  KKLT and K\"ahler uplifting scenarios,  sections ~\ref{11169_sec}, and~\ref{11226_sec}  look into the 2-parameter `swiss-cheese' CY $\mathbb{CP}_{11169}^4[18]$ in the LVS and K\"ahler uplifting scenario, and an anisotropic 3-parameter case based on the fibered CY $\mathbb{CP}_{11226}^4[12]$ relevant e.g. for the model of fibre inflation~\cite{Cicoli:2008gp}, respectively.


In all cases, we demand the relative magnitude $\Delta\equiv \delta V / V_0$ of the correction in the scalar potential $\delta V$ compared to the scalar potential $V_0$ employed in the given stabilisation mechanism to be somewhat small $\Delta < 1$, typically $\Delta \lesssim 0.1$ to guarantee survival of the original minimum of the stabilisation mechanism. This leads to an \emph{upper} bound $|W_0|\lesssim 10^{-3}$ on the flux superpotential in KKLT, and a \emph{lower} bound $\vo\gtrsim 10^8\ldots 10^9$ on the CY volume in the most generic version of LVS with ED3-brane instantons stabilizing the blow-up K\"ahler moduli for the minima to persist in presence of the GSW correction. The latter bound implies in particular, that the models of K\"ahler moduli inflation~\cite{Conlon:2005jm,Westphal:2005yz} including fibre inflation~\cite{Cicoli:2008gp} are under serious pressure from the new ${\cal O}(\alpha'^2)$-correction, since the bound $\vo > 10^8$ is difficult to reconcile with COBE normalization of the 
curvature perturbation and maintaining slow-roll flatness.
The method of K\"ahler uplifting does survive the presence of the new correction, if we stabilise the volume at $\vo \gtrsim 10^3$. This requires either a distastefully large-rank gauge groups on a D7-brane stack~\cite{Rummel:2011cd}, or a double-gaugino condensate racetrack sector for the positive-self-intersection K\"ahler modulus~\cite{Kallosh:2004yh,Sumitomo:2013vla}.

We expect our constraints to generalize to all type IIB CY flux compactifications except those where it depends just on a combination of the small blow-up volume moduli for which we give a criterion in Section~\ref{sec:GS}. Finally, all statements regarding moduli stabilisation in this paper do not take into account other $\alpha'^2$ corrections to the effective action as to our knowledge the GSW correction is the only one derived to this day. It is not excluded that other $\alpha'^2$ corrections appear even with the same moduli dependence. Hence, the results presented in this work provide a motivation to derive further $\alpha'$ corrections to the effective action of type IIB string theory/F-theory.



\section{Perturbations to no-scale supergravity}\label{sec:ex-no-scale}

In this section, we review the protection of the type IIB scalar potential by extended no-scale structure, following the original work of Cicoli, Conlon and Quevedo \cite{Cicoli:2007xp}. This extended no-scale structure has been shown to play a crucial role in the radiative stability of LVS, by ensuring that string loop corrections are subleading under some natural assumptions. At the heart of this approach lies a series expansion of the potential of the K\"ahler moduli sector in terms of small perturbations with respect to the leading behaviour, determined by the tree level K\"ahler potential $K_0$ and the flux superpotential $W_0$. We start by assuming that the 4 dimensional theory admits an expansion of the form
\be
K=K_0+\delta K\qquad\text{and}\qquad W=W_0+\delta W,
\label{eq:KandW}
\ee
where $\delta K$ represents the higher order perturbative corrections to the K\"ahler potential and $\delta W$ the nonperturbative corrections to the superpotential.

We want to compute the scalar potential
\be
V=e^K(F_a \overline{F}^a-3|W|^2)\,,
\label{eq:VF}
\ee
as a perturbative expansion in the small parameters $\delta K$ and $\delta W$. For our purposes it suffices to go up to second order in $\delta K$. Recalling that the correct K\"ahler coordinates are $T_a=\tau_a+i b_a$, the F-terms are given by
\be
F_a\equiv\frac{\partial W }{\partial T_a}+W \frac{\partial K}{\partial T_a}=\frac{\partial W }{\partial \tau_a}+\frac{W}{2} \frac{\partial K}{\partial \tau_a}\equiv W_a + \frac{W}{2} K_a,
\ee
where we have made use of the fact that $W=W(T)$ and that $K=K(T+\overline{T})$. The F-terms are contracted with the inverse K\"ahler metric
\be
K^{T_a\overline{T_b}}=\left(\frac{\partial^2 K}{\partial T_a \partial \overline{T_b}}\right)^{-1}= 4 \left(\frac{\partial^2 K}{\partial \tau_a \partial \tau_b}\right)^{-1}\equiv 4 K^{ab}.
\ee
In what follows we will work mostly with the real coordinates $\tau_a$ and $b_a$. Before expanding the F-term potential it is useful to recall some basic definitions and identities regarding the tree level K\"ahler potential
\be
K_0=-2 \log \vo.
\label{eq:K0}
\ee
The volume of the compactifications manifold is given in terms of the two cycle volumes $t_i$ and the triple intersection numbers $\kappa_{ijk}$ as
\be
\vo=\frac{1}{6}\kappa_{ijk}t^i t^j t^k.\label{eq:Volgen}
\ee
As customary one defines the four-cycle volumes as 
\be
\tau_i =\frac{\partial \vo}{\partial t^i}=\frac{1}{2}\kappa_{i j k}t^j t^k=\frac{1}{2} A_{i j} t^j\,,\label{eq:24cycles}
\ee
where the matrix $A$ is given by
\be
A_{i j}=\frac{\partial \tau_i}{\partial t^j}=\kappa_{ijk} t^k.
\ee
The tree level K\"ahler metric is 
\be
K^0_{ij}\equiv\frac{\partial^2 K}{\partial \tau_i\partial \tau_j}=\frac{t^i t^j}{2 \vo^2}-\frac{A^{-1}_{ij}}{\vo}\,,
\ee
and its inverse is
\be
K_0^{ij}=\tau_i\tau_j -\vo A_{ij}.
\label{eq:K0inverse}
\ee

One can show using the previous definitions that the K\"ahler potential of Eq. (\ref{eq:K0}) satisfies the following identities
\be
K^0_i K_0^{i j}=-\tau_i\qquad\text{and}\qquad K^0_i K_0^{i j} K^0_j =3. 
\label{eq:noscale}
\ee

The full metric is given by
\be
K_{ij}\equiv\frac{\partial^2 K }{\partial \tau_i\partial \tau_j}=K^0_{i j}+\delta K_{i j}\,,
\ee
and its inverse can be computed explicitly as a power series in derivatives of $\delta K$. By imposing $K_{ij} K^{jk}=\delta_i^k$ one finds
\be
K^{i j}=K_0^{ij}-K_0^{im}\delta K_{m n}K_0^{nj}+K_0^{im}\delta K_{m n}K_0^{np} \delta K_{p q}K_0^{qj}\,,
\ee
to second order in $\delta K_{i j}$.
We note that due to the expansion of the K\"ahler potential in Eq. (\ref{eq:KandW}) one may also write
\be
e^K=e^K\big|_0+e^K\big|_1+e^K\big|_2+\mc{O}(\delta^3)=\frac{1}{\vo^2}+e^K\big|_1+e^K\big|_2+\mc{O}(\delta^3)\,,
\ee
where the index $0,1,2,...,n$ denotes the order in the $\delta$ expansion. One can then expand the scalar potential of Eq. (\ref{eq:VF}) as 
\be
V=V_{(0,0)}+\delta V_{(1,0)}+\delta V_{(0,1)}+\delta V_{(1,1)}+\delta V_{(2,0)}+\delta V_{(0,2)}+\delta V_{(2,1)}+\mc{O}(\delta^3).
\label{eq:Vexpanded}
\ee
In the remaining of this paper we denote the terms in the scalar potential that involve m powers of $\delta K$ and n powers of $\delta W$ and/or their derivatives  by $\delta V_{(m,n)}$.

The leading order term, derived from the tree level K\"ahler potential and superpotential is given by
\be
V_{(0,0)}= e^K\big|_0 |W_0|^2 \left(K_a^0 K_0^{ab} K_b^0-3\right)=0.
\ee
and vanishes due to no-scale structure~\cite{Cremmer:1983bf,Ellis:1983sf}, Eq. (\ref{eq:noscale}).

The subleading correction  proportional to $\delta K$ and its derivatives is given by
\be
\delta V_{(1,0)}\equiv - e^K\big|_0 |W_0|^2 \left\{-2 K^0_a K_0^{ab} \delta K_b+K^0_a K_0^{am}\delta K_{mn} K_0^{nb} K^0_b\right\} +\frac{e^K\big|_1}{e^K\big|_0} V_{(0,0)},
\label{eq:dV10}
\ee
Noting that no-scale structure guarantees that the last term always vanishes and making use of the first identity in Eq. (\ref{eq:noscale}) one can simplify Eq. (\ref{eq:dV10}) to
\be
\delta V_{(1,0)}=-\frac{|W_0|^2}{\vo^2}\left\{2 \tau_b \delta K_b +\tau_m\delta K_{mn} \tau_n\right\}.
\label{eq:dV10a}
\ee
This is the leading correction to the scalar potential which is important both in the realisation of the LVS and K\"ahler uplifting, when $\delta K$ originates from the ten dimensional $ \alpha'^3\mc{R}^4$ term, and in the discussion of extended no-scale structure. We will analyse it in more detail in Section \ref{sec:extendedNS}.

The remaining terms in Eq. (\ref{eq:Vexpanded}) are given explicitly in Appendix \ref{Eqs}, which will be used to compute the corrections to the scalar potential originating from the GSW correction. Before proceeding we note that the first and second order corrections depending exclusively on $\delta W$, $\delta V_{(0,1)}$ and $\delta V_{(0,2)}$ respectively, give rise to two of the well known terms in the LVS potential. Assuming that the non-perturbative superpotential takes the form
\be
\delta W= \sum_j A_j e^{-a_j( \tau_j + i b_j) },
\ee
after axion minimisation one finds
\be
\delta V_{(0,1)} \sim - \frac{W_0 \overline{\delta W}_a+ \overline{W_0}\delta W_a}{\vo^2} \tau_a\sim - \frac{|W_0| A_s a_s \tau_s e^{-a_s \tau_s}}{\vo^2},
\ee
where we have used the fact that in `swiss-cheese' geometries the dominant non-perturbative effect is generated by the smallest cycles $\tau_s$. Considering that $ K_0^{s s}\sim \sqrt{\tau_s} \vo$,  one can show that
\be
\delta V_{(0,2)}\sim\frac{1}{\vo^2}\overline{\delta W}_a K_0^{ab}\delta W_b\sim \frac{ A_s^2 a_s^2 e^{-2 a_s \tau_s} K_0^{s s}}{\vo^2}\sim \frac{ A_s^2 a_s^2 e^{-2 a_s \tau_s}\sqrt{\tau_s}}{\vo},
\ee
to leading order in he $1/\vo$ expansion.

 Since this work focuses on the phenomenological effects of corrections to K\"ahler potential we will leave aside purely superpotential perturbations as these and analyse in detail corrections originating from $\delta K$ next.



\subsection{Moduli stabilisation and extended no-scale structure\label{sec:extendedNS}}

The proposal of any moduli stabilisation scenario must be accompanied by a study of its stability against higher order corrections. This was precisely what was done for the LVS in \cite{Cicoli:2007xp}. Building up on \cite{Berg:2007wt}, Cicoli, Conlon and Quevedo \cite{Cicoli:2007xp} have considered the impact of $g_s^2 \alpha'^2 $ corrections arising from string exchange between stacks of D3- and D7-branes on the LVS minimum. They have found that the combination of the extra $g_s $ suppression with the particular functional form of the corrections $\delta K$ ensured that the corrections to the LVS potential are subleading, ensuring this way that the LVS minimum would still exist. This was dubbed extended no-scale structure which we will now briefly review, focusing on the $\delta V_{(1,0)}$ term.
 
Assuming that the correction to the K\"ahler potential, $\delta K$, is an homogeneous function of degree n in the two cycle volumes, that is:
\be
t^a \frac{\partial \delta K}{\partial t^a}=n\delta K,
\ee
noting that 
\be
\tau_a \frac{\partial }{\partial \tau_a}=\frac{1}{2}t^a \frac{\partial }{\partial t^a}\,,
\ee
and that
\be
\tau_a\tau_b \frac{\partial^2 }{\partial \tau_a \partial \tau_b}=-\frac{1}{4}t^a\frac{\partial }{\partial t^a}+\frac{1}{4}t^a t^b \frac{\partial^2}{\partial t^a \partial t^b},
\ee
we can write Eq. (\ref{eq:dV10a}) as 
\be
\delta V_{(1,0)}=-\frac{|W_0|^2}{4 \vo^2}n (n+2)\delta K.
\label{eq:dV10b}
\ee
It is then clear that if the degree of $\delta K$ is -2, the correction to the scalar potential vanishes identically. This is what is understood by extended no-scale structure. The leading string loop corrections to K analysed in \cite{Cicoli:2007xp} feature this structure and therefore do not contribute to the scalar potential even though they might dominate the K\"ahler potential. 

This situation is to be contrasted with the $\alpha'^3$ correction of \cite{Becker:2002nn}, which plays a pivotal role in the LVS and in K\"ahler uplifting \cite{Balasubramanian:2004uy}. The $\alpha'^3$ corrected K\"ahler potential takes the form
\be
K=-2\log\left(\vo+\frac{\hat{\xi}}{2}\right)\sim \underbrace{{-2\log \vo} }_{K_0}\underbrace{-\hat{\xi}/\vo}_{\delta K}+\mc{O}(\vo^{-2}).
\label{eq:KLVS}
\ee
Clearly in this case $\delta K$ is not a degree -2, in fact one can show that it is degree -3 in the $t$'s and therefore it does generate a term in the scalar potential at this order. From Eq. (\ref{eq:dV10b}) setting $n=-3$ we find
\be
\delta V_{(1,0)}=-\frac{3|W_0|^2}{4\vo^2}\delta K=\frac{3|W_0|^2}{4\vo^3}\hat{\xi}.
\ee

Extended no-scale lends a level of protection in particular to moduli stabilisation scenarios that rely on the $\alpha'^3$ correction of \cite{Becker:2002nn} against radiative corrections to K. It assumes however that the numerical prefactors are of order unity and in certain cases it also relies on suppression by extra factors of $g_s$. This seems justified in the context of \cite{Cicoli:2007xp} but care is needed when dealing with corrections to $K$ that break these assumptions. In the next section we analyse the GSW correction in the same light and show how t can impose stringent limits on the various moduli stabilisation scenarios.

\section{$\alpha^{'2}$ corrected action}\label{sec:GS}

The quantum corrected volume of the CY threefold $X_3$ is~\cite{Grimm:2013gma}
\be
\tilde{\vo}=\vo-\frac{5 }{64}\vo_{D7\cap O7}\,,
\ee
and so the K\"ahler potential for the moduli becomes
\be
K=-2\log \tilde{\vo}=-2\log\left( \vo-\frac{5 }{64}\vo_{D7\cap O7} \right).
\ee
The correction is given in terms of the geometric quantity
\begin{equation}
 \vo_{D7 \cap O7} = 8 \int_{X_3} c_1(B_3)^2 \wedge J_b\,,\label{eq:c12J}
\end{equation}
where $c_1(B_3)$ and $J_b$ are the 1-st Chern class respectively K\"ahler form of the base $B_3$. The CY $X_3$ is constructed as the double-cover of $B_3$. This correction is derived in~\cite{Grimm:2013gma} from a higher derivative correction to the low energy limit of M-theory which is transported to F-theory using the duality between the two theories. Eq.~\eqref{eq:c12J} is obtained by taking the weak coupling (Sen) limit~\cite{Sen:1997gv} of F-theory, which yields type IIB string theory on $X_3$ with and O7 plane and D7-branes. The D7-branes have to be included to cancel the D7 tadpole, induced by the O7 plane which is $-8 c_1(B_3)$. In~\cite{Grimm:2013gma}, it is assumed for simplicity that the D7 tadpole is canceled by a single D7-brane with the characteristic Whitney umbrella shape~\cite{Collinucci:2008pf,Braun:2008ua} such that there are no non-Abelian singularities in the F-theory picture. For the following estimation of the relevance of the correction in the different moduli stabilisation scenarios, 
we will assume for simplicity that it takes the same form as given in Eq.~\eqref{eq:c12J} for the 
case that 
non-Abelian singularities are present.\footnote{While we expect the correction to have the same moduli dependence in the non-Abelian case, the numerical prefactor might change depending on the number of D7 branes. We thank Thomas Grimm for helpful discussions on this point.} This corresponds to D7-brane stacks in the IIB picture that can induce non-perturbative effects via gaugino condensation. 

Let us discuss the general dependence of Eq.~\eqref{eq:c12J} on the K\"ahler moduli in the LVS by considering $X_3$ to be of the `swiss-cheese' type. Let us for simplicity assume that there is only one large 4-cycle $\tau_1$ i.e., Eq.~\eqref{eq:Volgen} is given as
\begin{align}
 \begin{aligned}
 \vo &=\frac{1}{6} \left( \kappa_{111} (t^1)^{3} + \sum_{i,j,k=2}^{h^{1,1}} \kappa_{ijk} t^i t^j t^k \right)
     &= \gamma_1 \tau_1^{3/2} - \vo_s(\tau_2,..,\tau_{h^{1,1}})\,,\label{eq:swisscheese}
 \end{aligned}
\end{align}
with $\gamma_1 = \frac{\sqrt{2}}{3\sqrt{\kappa_{111}}}$ and $\vo_s(\tau_2,..,\tau_{h^{1,1}})$ is the small contribution to the overall volume $\vo$ that depends on the small 4-cycles $\tau_2,..,\tau_{h^{1,1}}$. In case of a `strong swiss-cheese' manifold $X_3$ this part is given as
\begin{equation}
 \vo_s(\tau_2,..,\tau_{h^{1,1}}) = \sum_{i=2}^{h^{1,1}} \gamma_i \tau_i^{3/2}\,,
\end{equation}
with $\gamma_i = \frac{\sqrt{2}}{3\sqrt{\kappa_{iii}}}$ and $\kappa_{ijk}=0$ unless $i=j=k$. As we will discuss later, there can be significant corrections to the scalar potential if Eq.~\eqref{eq:c12J} depends on $\tau_1$. For this reason, let us derive a condition for the vanishing dependence on $\tau_1$ for the case that $X_3$ is a toric CY whose volume can be written as in Eq.~\eqref{eq:swisscheese}. These CYs have been classified in~\cite{Kreuzer:2000xy} while attempts to find all `swiss-cheese' in the dataset of~\cite{Kreuzer:2000xy} can be found in~\cite{Gray:2012jy}.

Let $X_3$ be given by as a hypersurface equation in an ambient 4 complex dimensional space $X_4^{\text{amb}}$ which is a projective space $\mathbb{CP}^4$. Furthermore, let the base $B_3$ of the F-theory lift be given as a projective space $\mathbb{CP}^3$ as a subspace of $X_4^{\text{amb}}$. Generically $\text{dim}\,H^{1,1}(B_3,\mathbb{Z}) = \text{dim}\,H^{1,1}(X_3,\mathbb{Z}) \equiv h^{1,1}$ unless the CY hypersurface splits or does not intersect divisors of $X_4^{\text{amb}}$. In this case, $B_3$ is completely determined by a weight matrix $n_{ij} \in \mathbb{N}^{h^{1,1}\times h^{1,1}+3}$. Then,
\begin{equation}
 c_1(B_3) = \sum_i^{h^{1,1}+3} D_i = \sum_i^{h^{1,1}} k_i^b D^b_i\,,\label{c1B3toric}
\end{equation}
where $D_i$ are the toric divisors of $B_3$ and in the second equality in Eq.~\eqref{c1B3toric} we have chosen a basis $\{ \hat D^b_i\}$ of $H^{1,1}(B_3,\mathbb{Z})$. $\{D^b_i\}$ are the Poincar\'e dual 4-cycles to this basis. The coefficients $k_i^b\in \mathbb{Z}$ arise from the fact that only $h^{1,1}$ toric divisors are linear independent, i.e., there are three linear relations between the $h^{1,1}+3$ toric divisors which can be read off the weight matrix $n_{ij}$.\footnote{Consider for example the base $\mathbb{CP}_{1111}$. Since $D_1 = D_2 = D_3 = D_4 \equiv D$ we have $c_1(B_3)=D_1+D_2+D_3+D_4 = 4 D$.} The triple intersections $\kappa_{ijk}^b$ of the base divisors $\{D^b_i\}$ determine the volume in terms of the 2-cycle volumes $t_i^b$ according to Eq.~\eqref{eq:Volgen}. If $X_3$ is a `swiss-cheese' manifold, there exists a basis rotation $R_{ij}$ of the 4-cycles such that in the $\tau_i = R_{ij} \tau_j^b$, $\vo$ obtains the form given in Eq.~\eqref{eq:swisscheese}. For specific toric examples, the 
matrix $R_{ij}$ can be calculated according to the algorithm presented in~\cite{MayrhoferThesis,Gray:2012jy}. In this new basis $D_i = R_{ij} D^b_j$ we have
\begin{equation}
 c_1(B_3) = \sum_i^{h^{1,1}} k_i D_i \qquad \text{with} \qquad k_i \equiv R^{-1}_{ij} k_j^b \in \mathbb{Q}\,.
\end{equation}
The $\alpha'^2$ correction Eq.~\eqref{eq:c12J} is then given as
\begin{align}
 \begin{aligned}
  \vo_{D7 \cap O7} = 8 \left(\kappa_{111} k_1^2 t_1 + \sum_{i,j,k=2}^{h^{1,1}} \kappa_{ijk} k_i k_j t_k \right) = \frac{16}{3 \gamma_1} k_1^2 \sqrt{\tau_1} + \vo_{D7 \cap O7}(\tau_2,..,\tau_{h^{1,1}})\,,\label{voD7O7onebig}
 \end{aligned}
\end{align}
where $\vo_{D7 \cap O7}(\tau_2,..,\tau_{h^{1,1}})$ is the contribution to $\vo_{D7 \cap O7}$ that does not depend on $t_1$ and hence neither on $\tau_1$. For large $\gamma_1$, i.e., small self-intersection $\kappa_{111}$ of the large 4-cycle, $\vo_{D7 \cap O7}$ is suppressed w.r.t.\ the classical volume $\vo$. Furthermore, we see that the correction does not depend on $\tau_1$ iff $k_1=0$. If we would include $n_b$ large 4-cycles with positive intersection signature in Eq.~\eqref{eq:swisscheese}, this straightforwardly generalises to $k_i = 0$ for $i=1,..,n_b$. For a given toric CY, the condition $k_i=0$ can be easily checked once the transformation matrix $R_{ij}$ has been calculated.

Generically, $k_i$ and $\gamma_i$ can easily obtain values that are $\mathcal{O}(1..10)$ and $\mathcal{O}(10^{-2}..10^{-1})$ as we will also see in the examples in Sections~\ref{11169_sec} and~\ref{11226_sec}. As a consequence,
\begin{equation}
 \tilde \vo \sim \vo - \mathcal{O}(1..10) \sqrt{\tau_1}\,.\label{largefaccorr}
\end{equation}
we will see in the following that the $\alpha'^2$ correction can have a strong significance in the LVS and in K\"ahler uplifting even in the presence of the extended no-scale structure. Note that consistency of the low energy effective action requires the quantum corrected volume, $\tilde{\vo}$, to be positive definite. This translates into the following bound \footnote{ Higher order $\alpha'$ terms can in principle be used to relax this condition.}
\be
\frac{\vo_{D7\cap O7}}{\vo}<\frac{64}{5}\sim \mc{O}(10)\,,
\ee
at order $\alpha'^2$. 

Further assuming $\frac{\vo_{D7\cap O7}}{\vo}\ll \mc{O}(10)$ one can perform a series expansion in powers of $\frac{\vo_{D7\cap O7}}{\vo}$ and approximate
\be
K\sim-2\log\vo +\frac{5}{32} \frac{\vo_{D7\cap O7}}{\vo}+\left(\frac{5}{64} \frac{\vo_{D7\cap O7}}{\vo}\right)^2+\mathcal{O}\left( \frac{\vo_{D7\cap O7}}{\vo}\right)^3.\label{Kexpansion}
\ee
For convenience one defines $\delta K^{(1)}$ and  $\delta K^{(2)}$ to be the first and second order corrections to $K$ in the $\frac{\vo_{D7\cap O7}}{\vo}$ expansion, i.e.,
\be
\delta K^{(1)}\equiv \frac{5 }{32} \frac{\vo_{D7\cap O7}}{\vo}  \qquad,\qquad \delta K^{(2)}\equiv \left(\frac{5}{64} \frac{\vo_{D7\cap O7}}{\vo}\right)^2=\left(\frac{\delta K^{(1)}}{2}\right)^2.
\ee
For the case of one big 4-cycle $\tau_1$ in a `swiss-cheese' CY this can be approximated as
\begin{equation}
 \delta K^{(1)} \simeq \frac{5 k_1^2}{6 \gamma_1^2 \tau_1} \simeq \frac{5 k_1^2}{6\gamma_1^{4/3} \vo^{2/3}}\,,\label{deltaKswiss}
\end{equation}
using Eq.~\eqref{voD7O7onebig} and $\vo \simeq \gamma_1 \tau_1^{3/2}$. Also note that one may perform the following expansion
\be
e^K= \left(\vo-\frac{5}{64}\vo_{D7\cap O7}\right)^{-2}=\underbrace{\frac{1}{\vo^2} }_{\equiv e^K\big|_0}+ \underbrace{\frac{5}{32}\frac{\vo_{D7\cap O7}}{\vo^3}}_{\equiv e^K\big|_1}+\underbrace{\frac{75}{4096}\left(\frac{\vo_{D7\cap O7}}{\vo^2}\right)^2}_{\equiv e^K\big|_2}+ \mathcal{O}\left( \frac{\vo_{D7\cap O7}}{\vo}\right)^3.
\ee

The volume of the intersection between the D7-brane and the O7 plane is parametrized by the two cycle volumes or alternatively by some function of degree $1/2$ in the four cycle volumes $\tau_a$. Let us assume that 
\be
\sum_a\tau_a \frac{\partial \vo_{D7\cap O7}}{\partial \tau_a}=m  \vo_{D7\cap O7}.
\label{eq:corr1}
\ee
This corresponds to stating that the intersection volume is a homogeneous function of degree $2m$ in the two cycle volumes. One can then show that 
\be
\tau_a \delta K^{(1)}_{a b}=\left(m-\frac{5}{2}\right) \delta K^{(1)}_b\qquad \text{and}\qquad \tau_a \delta K^{(2)}_{a b}=2 \left(m-2\right) \delta K^{(2)}_b.
\label{eq:corr2}
\ee
Equations (\ref{eq:corr1}) and (\ref{eq:corr2}) allow for a considerable simplification in the computation of the corrections to the scalar potential given by Eqs. (\ref{eq:dV10})-(\ref{eq:dV20}). For the cases under consideration $m=1/2$, which implies that $\delta K^{(1)}$ and $\delta K^{(2)}$ are homogeneous functions of degree $-2$  and $-4$ respectively in the two cycle volumes. Therefore this correction has the same form as that of \cite{Cicoli:2007xp} and so should naively be protected by the same extended no-scale structure arguments.

Noting that $\delta V_{(1,0)}$ is a linear in $\delta K$ we may write it as
\be
\begin{split}
\delta V_{(1,0)}(\delta K)&=\delta V_{(1,0)}(\delta K^{(1)})+\delta V_{(1,0)}(\delta K^{(2)})\\
&=\delta V_{(1,0)}(\delta K^{(2)})=-\frac{|W_0|^2}{2\vo^2}\left[\delta K^{(1)}\right]^2,\label{dV1gen}
\end{split}
\ee
where we have used the fact that $\delta V_{(1,0)}(\delta K^{(1)})=0$ since $\delta K^{(1)}$  is homogeneous of degree $-2$ in the t's. Since $\delta K^{(1)}\sim \vo^{-2}$ we expect either
\be
\delta V_{(1,0)}\sim \vo^{-4} \qquad\text{or}\qquad\delta V_{(1,0)}\sim\vo^{-10/3},
\ee
depending whether or not $\vo_{D7\cap O7}$ is proportional to $\vo^{1/3}$. Both possibilities are more volume suppressed than the LVS/K\"ahler uplifting potential, which is proportional to $\vo^{-3}$, so in principle they can be made subleading by considering large enough values for $\vo$.

The $(2,0)$ component of the scalar potential admits an expansion of the form
\be
\begin{split}
\delta V_{(2,0)}(\delta K)&=\delta V_{(2,0)}(\delta K^{(1)}+\delta K^{(2)})=\delta V_{(2,0)}(\delta K^{(1)})+\mc{O}\left(\frac{\vo_{D7\cap O7}}{\vo}\right)^3\\
&=\frac{|W_0|^2}{\vo^2}\delta K^{(1)}_a K_0^{ab}\delta K^{(1)}_b+\mc{O}\left(\frac{\vo_{D7\cap O7}}{\vo}\right)^3.\label{dV2gen}
\end{split}
\ee
Note that $\delta K^{(2)}$ also contributes to $\delta V_{(2,0)}$, however we can consistently neglect it due to extra volume suppression.  The same observation holds true for the $\delta V_{(1,1)}$ and $\delta V_{(1,2)}$ terms: the dominant contributions to both $\delta V_{(1,1)}$ and $\delta V_{(1,2)}$ will come from $\delta K^{(1)}$ as these will be the less volume suppressed. In order to determine the typical size of these terms let us once again consider the `swiss-cheese' case where $\delta K^{(1)}$ is given in Eq.~\eqref{deltaKswiss}. Using the general formulas in Eq.~\eqref{dV1gen} and Eq.~\eqref{dV2gen} we find
\begin{equation}
 \delta V_{(1,0)} + \delta V_{(2,0)} \simeq \frac{25}{216}\,\frac{|W_0|^2 k_1^4}{\gamma_1^{6}\, \tau^5} \simeq \frac{25}{216}\,\frac{|W_0|^2 k_1^4}{\gamma_1^{8/3} \vo^{10/3}}\,.\label{dVswiss}
\end{equation}
We will discuss the typical size of these corrections using three representative examples in the next section.

\subsection{Example 1: Single modulus case}\label{sec:KKLT}

Let us start by looking at the single modulus case put forward in \cite{Grimm:2013gma}. While it is not possible to realise the large volume scenario in such a compact space, since it has a single K\"ahler modulus, this is the simplest geometry in which one can build the KKLT~\cite{Kachru:2003aw} and K\"ahler uplifting~\cite{Westphal:2006tn,Rummel:2011cd} scenarios. We will therefore investigate how the $\alpha'^2$ correction to the action affects the minimum in these two scenarios.

The starting point for this analysis is the $\alpha'^2$ corrected K\"ahler potential  \cite{Grimm:2013gma}
\be
K=-\log\left[\frac{2}{9} \tau\left(\tau-\frac{15}{8} k^2\right)^2\right],\label{Ksingle}
\ee
this follows from Eq.~\eqref{voD7O7onebig} where $k\equiv k_1$ is the first Chern number of the base manifold $B_3$ and $\gamma_1 = \sqrt{2}/3$ using $\kappa_{111}=1$. Positivity of the quantum corrected volume requires $\tau>\frac{15}{8} k^2$.

\subsubsection*{KKLT}

We assume that the volume modulus is stabilised by the KKLT mechanism, where the superpotential is the sum of a flux term and a nonperturbative term: 
\be
W=W_0+A e^{-a \tau}.
\ee
In this context, the largeness of $\tau$ is made possible by the smallness of the flux superpotential $W_0$. Recalling that before uplifting, the KKLT minimum is SUSY AdS, we find
\be
D_T W=0\Leftrightarrow W_0=- A e^{-a \langle\tau\rangle}\left(1+2 a\langle\tau\rangle\frac{1-\frac{8\langle\tau\rangle}{15 k^2}}{1-\frac{24\langle\tau\rangle}{15 k^2}}\right)\sim-A e^{-a \langle\tau\rangle},
\label{eq:W0SUSY}
\ee
and so the flux superpotential must lie in the interval
\be
-A e^{-a 15 k^2/8}\lsim W_0<0.
\ee
As an illustration, if one sets $A=1,\:a=0.1$ this bound becomes $\mc{O}(-10^{-2})<W_0<0$ for $k=4$. We can then see that the large values of $\tau$ required for consistency of the action to order $\alpha'^2$ impose a tuning of the flux superpotential. 
In regions of moduli space where $\tau\gg 15 k^2/8$ one can expand the potential in powers of $\vo_{D7\cap O7}/\vo$ and follow the method outlined above to show that the K\"ahler potential admits an expansion of the form $K\sim K_0+\delta K$ where
\be
\delta K=\underbrace{2 \left(\frac{15}{8}\right) \frac{k^2}{\tau}}_{\equiv \delta K^{(1)}}+\underbrace{\left(\frac{15}{8}\right)^2\frac{k^4}{\tau^2}}_{\equiv \delta K^{(2)}}\,,\label{deltaKonemod}
\ee
obtained from Eq.~\eqref{deltaKswiss} and $K_0=-2\log \vo=-\log (2/9 \tau^3)$. The scalar potential is then given by
\be
V=V_{KKLT}+\delta V_{(1,0)}+ \delta V_{(2,0)}+\delta V_{(1,1)}+\delta V_{(1,2)}+\mc{O}\left(\frac{\vo_{D7\cap O7}}{\vo}\right)^3,
\label{eq:VtotKKLT}
\ee
where the KKLT potential is
\be
 V_{KKLT}\equiv \delta V_{(0,1)}+ \delta V_{(0,2)}=\frac{18 a A^2 e^{-2 a \tau }}{\tau ^2}+\frac{18 a A e^{-a \tau } W_0}{\tau ^2}+\frac{ 6 a^2 A^2 e^{-2 a \tau }}{\tau },
\ee
and the leading $\alpha'^2$ corrections Eq.~\eqref{dVswiss} take the form
\be
\delta V_{(1,0)}+\delta V_{(2,0)}=-\frac{675 k^4}{64}\frac{W_0^2}{\tau ^5}\,,\label{dV12onemod}
\ee
and
\be
\delta V_{(1,1)}+\delta V_{(1,2)}=45 k^2  \frac{ A^2 a e^{-2 a \tau }}{\tau ^3}+45 k^2  \frac{W_0 A a e^{-a  \tau } }{\tau ^3}+15 k^2  \frac{ A^2 a^2 e^{-2 a \tau }}{2 \tau ^2}.
\ee

Assuming one accepts the level of tuning of $W_0$ required to achieve $\tau\gg 15 k^2/8$, one may ask if the KKLT minimum survives at the $\alpha'^2$ level. To answer this question one must compare the typical size of the several contributions to the full potential, Eq. (\ref{eq:VtotKKLT}), at the putative minimum. Making use of Eq. (\ref{eq:W0SUSY}) we see that at the KKLT minimum
\be
\langle V_{KKLT}\rangle=-\frac{6 A^2 a^2 e^{-2 a  \tau }  \left(15 k^2/4 -2 \tau \right)^2}{\left(5 k^2/4  -2 \tau \right)^2 \tau }\,.
\ee
Moreover one can show that that the $\alpha'^2$ corrections are small, provided $\langle\tau\rangle$ is sufficiently large:
\be
\left\langle\frac{\delta V_{(1,0)}+\delta V_{(2,0)}}{V_{KKLT}}\right\rangle \sim \frac{25}{32} \frac{k^4}{\langle\tau\rangle^2}\ll1
\qquad
\text{and}\qquad
\left\langle\frac{\delta V_{(1,1)}+\delta V_{(1,2)}}{V_{KKLT}}\right\rangle\sim \frac{15}{4} \frac{k^2}{\langle\tau\rangle}\ll1.
\ee

Clearly the mixed term $\delta V_{(1,1)}+\delta V_{(1,2)} \sim \delta V$ is the dominant contribution to the potential coming from the GSW correction. Taking for concreteness $k=4$ which is the CY $\mathbb{CP}_{11114}^4[8]$, we see that $\delta V$ is of the order of the KKLT potential at $\tau\sim 60$, which corresponds to tuning $W_0\sim -10^{-3}$\ \footnote{Care is needed in this small volume region, since even though the quantum corrected volume is positive, the series expansion in $\vo_{D7\cap O7}/\vo$ converges slowly and so for a definite result higher order terms should be taken into account.}. The hierarchy $\Delta\equiv\delta V/V_{KKLT}$ drops to $1/10$ at $\tau \sim 600$. In this regime one can assume that the KKLT minimum is preserved, however this comes at the cost of tuning the superpotential to $W_0\sim -10^{-25}$. For the GSW correction to give rise to a mere 1\% level correction to KKLT, we are pushed to higher volumes still, around $\tau\sim 6000$, which implies $W_0\sim -10^{-258}$. We note 
that this tuning in $W_0$ follows from the structure 
of the KKLT minimum and it is only more severe than the fiducial example in \cite{Kachru:2003aw} due to the requirement of significantly larger volumes. We are then led to the conclusion that for large $\langle\tau\rangle$  ({\it extremely} small $W_0$)  the usual KKLT minimum is preserved since the  $\alpha'^2$ corrections become subleading. Just preserving stability requires $\Delta\lesssim {\cal O}(1)$ which needs only moderate tuning $|W_0|\lesssim 10^{-3}$. However, getting additional suppression of the GSW correction to $\Delta\ll 1$ comes with an extreme exponential cost in the tuning of $W_0$ which will turn out to be much more severe than the corresponding changes in the CY volume in the LVS scenario to which we turn after the next subsection.

\subsubsection*{K\"ahler uplifting}

In the absence of the $\alpha'^2$ correction of Eq.~\eqref{eq:c12J} to the K\"ahler potential, the dominant terms in the scalar potential of the K\"ahler uplifting scenario arise from non-perturbative contributions from gaugino condensation on 7-brane stacks and the $\alpha'^3$-correction~\cite{Becker:2002nn,Rummel:2011cd}
\begin{equation}
 V \simeq \underbrace{\frac{2 |W_0|^2 \hat \xi}{4 \vo^3}}_{\equiv V_{\alpha'^3}} - \frac{2 W_0 a\,A\, e^{-a t}}{\gamma^{2/3} \vo^{4/3}}\,,
\end{equation}
where $V_{\alpha'^3}$ is a measure for the barrier height of the de Sitter minimum of the potential. The leading order terms induced by the $\alpha'^2$ correction are $\delta V_{(1,0)} + \delta V_{(2,0)} \sim \vo^{-10/3}$ given in Eq.~\eqref{dV12onemod}. We can estimate their significance by building the quotient
\begin{equation}
 \left| \frac{\delta V_{1,0}+\delta V_{2,0}}{V_{\alpha'^3}} \right| 
 \simeq  \frac{25}{12\cdot 6^{1/3}}\,\frac{k^4}{ \hat \xi\, \vo^{1/3}} = \frac{25}{12\cdot 6^{1/3}}\,\frac{k^4}{ \epsilon\, \vo^{4/3}}\,,\label{relpotKUP}
\end{equation}
where in the last equation we have inserted $\epsilon \equiv \hat \xi / \vo$ which is required to be small in the K\"ahler uplifting scenario~\cite{Rummel:2011cd}. Since $\hat \xi \propto g_s^{-3/2}$ this quantity can in principle be made arbitrarily large in the weak coupling limit $g_s \to 0$. For $k=4$ and $\epsilon = 0.1$ the $\delta V_{(1,0)} + \delta V_{(2,0)}$ terms become of the same order as $V_{\alpha'^3}$ for $\vo \simeq 400$ in string units. For this value of the volume the expansion coefficient in the K\"ahler potential Eq.~\eqref{deltaKonemod} is $\delta K^{(1)}\simeq 0.6$ such that the expansion converges rather slowly. Hence, we check numerically for a K\"ahler uplifted de Sitter vacuum of the full scalar potential including the $\alpha'^2$ correction. Our chosen parameters are presented in Table~\ref{KUpTab}.
\begin{table}[h!]
\centering
  \begin{tabular}{c|c|c|c|c}
  $W_0$ & $A$ & $a$ & $\hat \xi$ & $\langle \vo \rangle$\\
  \hline
  $-0.52$ & $1$ & $2\pi/110$ & $49.8$ & $347$\\
  \end{tabular}
  \caption{K\"ahler uplifted de Sitter vacuum obtained by numerical analysis of the full scalar potential. The volume is given in string units.}
  \label{KUpTab}
\end{table}
Our analytical and numerical analysis show that we do not have to demand for a large suppression of the $\delta V_{(1,0)} + \delta V_{(2,0)}$ terms w.r.t.\ to $V_{\alpha'^3}$. A de Sitter vacuum can still be found in the $\alpha'^2$ corrected scalar potential by choosing appropriate values for the parameters, in this case $W_0$, $A$, $a$ and $\hat \xi$. This is also true in the multi moduli `swiss-cheese' case for K\"ahler uplifting as we will discuss in the next Section.

Let us note again, that dS minima persist at moderate volumes $\vo\sim 400$ while refraining from obscenely large gauge group with rank much larger than ${\cal}(100)$ only for values of the expansion coefficient of the K\"ahler potential $\delta K^{(1)}\lesssim {\cal O}(1)$. In such a regime we cannot be confident that higher-order $\alpha'$-corrections might not enter the K\"ahler potential with an expansion coefficient of similar size. As this limits trust in the scalar potential derived in the above regime, one needs either a distastefully large-rank gauge group~\cite{Rummel:2011cd}, or a racetrack double-gaugino condensate for the K\"ahler modulus with positive self-intersection~\cite{Kallosh:2004yh,Sumitomo:2013vla} to push the volume into a regime where $\delta K^{(1)}\ll 1$.

\subsection{Example 2: $\mathbb{CP}_{11169}^4[18]$} \label{11169_sec}

We now focus on the simplest geometry that allows for a LVS minimum: the two modulus `swiss-cheese'. As an explicit realisation of this geometry  we consider the degree 18 hypersurface in complex projective space $\mathbb{CP}_{11169}^4[18]$ used in the original LVS construction \cite{Balasubramanian:2005zx}.
The classical volume of this manifold is 
\be
\vo=\frac{1}{18} \tau_1^{3/2}-\frac{1}{9} \sqrt{2} \tau_2^{3/2}\,,\label{Vol11169}
\ee
and the K\"ahler potential can be written in terms of the $\alpha'^2$ and $\alpha'^3$ corrected volume as
\be
K=-2 \log\left[\frac{1}{18} \tau_1^{3/2}-\frac{1}{9} \sqrt{2} \tau_2^{3/2}-\frac{135}{8}  \sqrt{\tau_1}+\frac{\hat{\xi}}{2}\right],\label{K11169}
\ee
which allows one to define the corrections to the tree level K\"ahler potential $K_0=-2\log \vo$ as
\be
\delta K^{(1)}=\frac{1215}{2}  \frac{\sqrt{\tau_1}}{\tau_1^{3/2}-2 \sqrt{2} \tau_2^{3/2}}\qquad \text{and}\qquad\delta K^{(2)}= \frac{1476225}{16} \left( \frac{\sqrt{\tau_1}}{\tau_1^{3/2}-2 \sqrt{2} \tau_2^{3/2}}\right)^2.
\ee
We note that there are also contributions to $\delta K$ from the $\alpha'^3$ term which were given in Eq. (\ref{eq:KLVS}). 

\subsubsection*{LVS}

We assume that the small modulus, $\tau_2$, supports the nonperturbative effect, giving rise to a superpotential of the form
\be
W=W_0+  A e^{-a \tau_2}.
\label{eq:WLVS}
\ee
The scalar potential is then given by 
\be
V=V_{LVS}+\delta V_{(1,0)}+\delta V_{(2,0)}+\delta V_{(1,1)}+\delta V_{(1,2)}+\mc{O}\left(\frac{\vo_{D7\cap O7}}{\vo}\right)^3,
\ee
where $V_{LVS}$ is the usual large volume scenario potential for $\mathbb{CP}_{11169}^4[18]$: 
\be
V_{LVS}=12 \sqrt{2} \frac{ a^2 A^2 e^{-2 a \tau_2} \sqrt{\tau_2}}{\vo }-4 \frac{ a A e^{-a \tau_2} W_0 \tau_2}{\vo ^2}+\frac{3}{4}\frac{W_0^2 \hat{\xi} }{\vo ^3}\,,
\label{eq:VLVS}
\ee
and the corrections take the form
\be
\delta V_{(1,0)}+\delta V_{(2,0)}=-\frac{18225}{16} \left(\frac{3}{2}\right)^{1/3}  \frac{W_0^2 }{\vo ^{10/3}}\,,
\label{eq:V10V20P11169}
\ee
and
\be
\delta V_{(1,1)}+\delta V_{(1,2)}=\frac{405}{2^{1/6}} 3^{2/3} \frac{A^2 a^2   \sqrt{\tau_2}\  e^{-2 a  \tau_2} }{\vo ^{5/3}}-135\ 2^{1/3} 3^{2/3}  \frac{  W_0   A a \tau_2 e^{-a \tau_2}  }{\vo ^{8/3}}.
\label{eq:V11P11169}
\ee

Clearly the corrections of Eqs. (\ref{eq:V10V20P11169}) and (\ref{eq:V11P11169}) are subleading in the volume, being respectively proportional to $\vo^{-10/3}$ and $(\log \vo)^{3/2}\vo^{-11/3}$ at the LVS minimum. We note that, as expected, these corrections have the same functional form as the string loop corrections of \cite{Cicoli:2007xp}. However the corrections of  \cite{Cicoli:2007xp} were of order $\alpha'^2 g_s^2 $ and had overall numerical coefficients of $\mc{O}(1)$, while here we find that {\it the overall coefficients in Eqs. (\ref{eq:V10V20P11169}) and (\ref{eq:V11P11169}) are neither $\mc{O}(1)$ nor feature suppression by extra powers of $g_s\ll1$.} Therefore the extended no-scale structure is of limited use and these $\alpha'^2$ terms in the scalar potential are only subleading at very large volumes. This restricts the range of validity of the LVS potential and its phenomenology in such a compact space.

In order quantify the effect of the GSW corrections on the minimum of Eq. (\ref{eq:VLVS}), let us recall that it is located at 
\be
\langle\vo\rangle=\frac{3}{\sqrt{2}} \frac{W_0}{a A}\sqrt{\langle\tau_2\rangle}e^{a\langle\tau_2\rangle}\left( 1-\frac{3}{4 a \langle\tau_2\rangle}\right) \qquad,\qquad\langle\tau_2\rangle^{3/2}=\frac{9 \sqrt{2}}{4}\hat{\xi}\left(1+\frac{1}{2 a \langle\tau_2\rangle}\right)\,,
\label{eq:minPos}
\ee
 to leading order in the $1/a \tau_2$ expansion. The depth of the AdS minimum is
\be
\langle V_{LVS}\rangle= -\frac{3}{8}\frac{|W_0|^2 \hat{\xi}}{\langle\vo\rangle^3\ a \langle\tau_2\rangle}.
\label{eq:LVSvev}
\ee
Using Eq. (\ref{eq:minPos}) once can further simplify this to
\be
\langle V_{LVS}\rangle \sim -\frac{3}{8}\frac{|W_0|^2 \hat{\xi}}{\langle\vo\rangle^3\ \log\langle\vo\rangle} \simeq-\frac{\sqrt{\log \langle\vo\rangle}}{6\sqrt{2}\,a^{3/2}\,\langle\vo\rangle^3}|W_0|^2 \,,
\label{eq:minPos2}
\ee
where we have eliminated the parameter $\hat{\xi}$ in favour of $\langle\tau_2\rangle^{3/2} \sim (\log \langle\vo\rangle)^{3/2} /a^{3/2}$ in the final step.  

Note that the least volume suppressed terms near the putative LVS minimum are the pure K\"ahler terms of Eq. (\ref{eq:V10V20P11169}) . We therefore evaluate the validity of the LVS minimum by looking at the ratio $\Delta\equiv\delta V/V_{LVS}$ where $\delta V=\delta V_{(1,0)}+\delta V_{(2,0)}$ and requiring it to be sufficiently small.  Taking the ratio between Eq (\ref{eq:V10V20P11169}) and Eq. (\ref{eq:LVSvev}) one finds at next-to-next-leading-logarithmic order that the volume must satisfy the following bound

\be
\langle\vo\rangle \gtrsim 2.9 \times 10^{15}\times \frac{\left(\frac{a}{2\pi }\right)^{9/2} \Delta^{-3}}{94+\frac{13}{2}\ln\left[\left(\frac{a}{2\pi }\right)^{3} \Delta^{-2}\right]}\,.
\ee

In Table \ref{tab:volumeBounds1} we present the lower limit on the volume of the compact space for different values of $a $ and $\Delta$.\footnote{The volume bound will depend on the intersection numbers $\gamma_1,k_1$ governing the classical volume and the correction, respectively, in a structurally similar way as it depends on $\Delta$. This is, because $\gamma_1,k_1$ enter the relevant terms of the scalar potential as power-law factors with ${\cal O}(1)$ exponents, see Eq.~\eqref{dVswiss}.}
\begin{table}[t!]
\centering
\begin{tabular}{l ||c|c}
&$a=2\pi$&$a=2\pi/10$\\
\hline
\hline
$\Delta=1$ & $\sim 10^{13}$ & $\sim 10^{9}$ \\
\hline
$\Delta=10^{-1}$ & $ \sim 10^{16}$ & $\sim 10^{12}$ \\
\hline
$\Delta=10^{-2}$ & $\sim 10^{19}$ & $\sim 10^{15}$ \\
\end{tabular}
\caption{Bounds on the volume for LVS on $\mathbb{CP}_{11169}^4[18]$. }
\label{tab:volumeBounds1}
\end{table}

As can be seen in Table \ref{tab:volumeBounds1}, the worst case scenario for LVS stability is the generic one with non-perturbative effects arising from ED3 branes, for which $a=2\pi$. Constraints on the volume are weaker for gaugino condensation on stacks of D7-branes. Nonetheless one sees that the regime of validity of the large volume scenario is pushed towards larger volumes than previously assumed by relying on the extended no-scale protection.

While there seems to be a limited margin for tuning particular models to evade  these constraints, one thing seems clear: the lower limits on $\vo$ put models of K\"ahler moduli inflation, such as those of \cite{Conlon:2005jm,Bond:2006nc}, under severe strain since they typically require considerably smaller volumes in order to give rise to the correct amplitude of density perturbations.

Finally, let us note here that our stability criterion $\Delta \lesssim 1$ can be rigorously justified by calculating the mass matrix of the canonically normalised moduli to leading order in the non-perturbative, $\alpha'^2$- and $\alpha'^3$-corrections. Demanding positivity of the eigenvalues of the Hessian amounts to requiring $\Delta\lesssim {\cal O}(2)$ which essentially justifies our analysis.

\subsubsection*{K\"ahler uplifting}

The one-modulus case discussed in Section~\ref{sec:KKLT} can be easily extended to the multi moduli `swiss-cheese' scenario of K\"ahler uplifting. This is for the simple reason that the GSW correction is only relevant if it is dominated by the large 4-cycle $\tau_1$. Hence, the generalization of Eq.~\eqref{relpotKUP} is
\begin{equation}
 \left| \frac{\delta V_{1,0}+\delta V_{2,0}}{V_{\alpha'^3}} \right| \simeq  \frac{25}{162}\,\frac{k_1^4}{\gamma_1^{8/3} \epsilon\, \vo^{4/3}}\,.\label{dVKUPrelgen}
\end{equation}
Eq.~\eqref{dVKUPrelgen} is $\lesssim 1$ if the classical volume fulfills the bound
\begin{equation}
 \vo \gtrsim \left(\frac{25}{162}\right)^{3/4} \frac{k_1^3}{\gamma_1^2 \epsilon^{3/4}} \simeq \frac{\kappa_1\, k_1^3}{\epsilon^{3/4}}\,.\label{volboundKUPgen}
\end{equation}
This implies
\begin{equation}
 \delta K^{(1)} \lesssim 3\sqrt{\frac{\epsilon}{2}}\,,
\end{equation}
using Eq.~\eqref{deltaKswiss} such that small $\epsilon$ or larger volumes than implied by Eq.~\eqref{volboundKUPgen} have to be taken into account in order for the expansion of the K\"ahler potential Eq.~\eqref{Kexpansion} to be justified.

In the two-parameter model $\mathbb{CP}_{11169}^4[18]$ under consideration we have a rather large self-intersection of $\kappa_1 = 72$ and $k_1 = 3/2$ such that for $\epsilon = 0.1$ the bound in Eq.~\eqref{volboundKUPgen} is given as $\vo \gtrsim 1400$. In~\cite{Louis:2012nb}, a fully consistent model of K\"ahler uplifting on $\mathbb{CP}_{11169}^4[18]$ has been constructed, before the GSW correction was derived. The classical volume of the CY in this construction is $\vo \simeq 52$ which is clearly a regime in which the GSW correction $\delta V_{1,0}+\delta V_{2,0}$ contributes very strongly to the scalar potential. If there would be no further corrections at the level of $\alpha'^2$ to the K\"ahler potential with the same moduli dependence this would put these explicit de Sitter models under severe strain. Together with the constraints on the LVS scenario discussed in this work, this provides strong motivation for studying further $\alpha'$ corrections to the IIB/F-theory effective action, as already 
mentioned in the introduction.

\subsection{Example 3: $\mathbb{CP}_{11226}^4[12]$} \label{11226_sec}

Some phenomenologically interesting constructions of the LVS rely on manifolds that are not of the `swiss-cheese' form but are instead fibered. In fibered LVS constructions the overall volume is not controlled by a single modulus like in the `swiss-cheese' case, being instead given by $\vo\sim \sqrt{\tau_1} \tau_2$ and therefore dependent on two K\"ahler moduli. 

The fiducial example of such a manifold is  $\mathbb{CP}_{11226}^4[12]$, which has inspired a multitude of phenomenological applications within the context of LVS: from the study of string loop corrections \cite{Cicoli:2007xp} to inflation \cite{Cicoli:2008gp,Cicoli:2011ct} and a stringy realisation of the supersymmetric large extra dimensions scenario \cite{Cicoli:2011yy,Cicoli:2012tz}. Strictly speaking the $\mathbb{CP}_{11226}^4[12]$ is a two parameter manifold, however one usually adds a blow-up by hand  to stabilise the overall volume {\it \`{a} la} LVS. The volume is then
\be
\vo=\frac{1}{6} \sqrt{\tau_1} (-2 \tau_1+3 \tau_2)-c_1 \tau_3^{3/2},
\ee
and so the K\"ahler potential can be written as 
\be
K=-2 \log\left[\frac{1}{6} \sqrt{\tau_1} (3 \tau_2 -2 \tau_1)-c_1 \tau_3^{3/2}-\frac{45 (2 \tau_1+\tau_2+c_2 \tau_3)}{8 \sqrt{\tau_1}}+\frac{\hat{\xi}}{2}\right]\,, \label{K11226}
\ee
to order $\alpha'^3$. Here we are using the constants $c_1$ and $c_2$ to parametrise our ignorance regarding the blow-up modulus extension of the compact $\mathbb{CP}_{11226}^4[12]$.  If one wants to have an LVS minimum then clearly $c_1\neq0$, however without an explicit geometrical construction we do not know whether the $\alpha'^2$ correction depends or not on $\tau_3$. In what follows we perform a generic analysis, leaving $c_2$ unspecified.  The $\alpha'^2$ corrections to the K\"ahler potential are
\be
\delta K^{(1)}=-\frac{135}{2} \frac{ 2 \tau_1+\tau_2+c_2 \tau_3}{2 \tau_1^2-3 \tau_1 \tau_2+6 c_1 \sqrt{\tau_1} \tau _3^{3/2}},
\ee
and
\be
\delta K^{(2)}=\frac{18225}{16} \left( \frac{ 2 \tau_1+\tau_2+c_2 \tau_3}{2 \tau_1^2-3 \tau_1 \tau_2+6 c_1 \sqrt{\tau_1} \tau _3^{3/2}} \right)^2.
\ee

Assuming that the nonperturbative effects originate from $\tau_3$, the superpotential takes the form
\be
W=W_0+A e^{-a \tau_3}.
\ee
One can then show that the leading order potential depends only on two of the original three directions in K\"ahler moduli space, taking the form
\be
V_{LVS}=\frac{8}{3}\frac{ a^2 A^2 e^{-2 a \tau_3} \sqrt{\tau_3}}{c_1 \vo }-4 \frac{ a A e^{-a \tau_3} W_0 \tau_3}{\vo ^2}+\frac{3}{4}\frac{W_0^2 \hat{\xi}}{\vo^3}.
\label{eq:VLVSfibre}
\ee
In writing Eq. (\ref{eq:VLVSfibre}) we have eliminated $\tau_2$ in terms of $\vo$ which leaves $V_{LVS}$ independent of $\tau_1$. The fact that at this level $\tau_1$ is a flat direction has proven an interesting feature of these geometries, giving rise to the above mentioned phenomenological applications. Usually the stabilisation of $\tau_1$ is achieved by the inclusion of subleading effects into the action: string loops and poly-instantons have both been used to this effect. Even though  it is conceivable that the $\alpha'^2$ corrections can play a similar role we will refrain from performing such analysis here and will instead focus on the study of the stability of the LVS once these corrections are included.

The subleading contributions to the scalar potential take the form
\be
\begin{split}
\delta V_{(1,0)}+\delta V_{(2,0)}=& \frac{2025}{8} \frac{W_0^2}{\vo ^2 \tau_1^2}-675 \frac{W_0^2}{\vo ^3 \sqrt{\tau_1}}+\frac{675}{8}c_2^2 \frac{ W_0^2 \sqrt{\tau_3}}{c_1 \vo ^3 \tau_1}\\
&+\frac{2025}{8} c_2 \frac{W_0^2 \tau_3}{\vo ^3 \tau_1^{3/2}}+\frac{2025}{8}  c_1\frac{W_0^2 \tau_3^{3/2}}{\vo ^3 \tau_1^2}\,,
\end{split}
\ee
and
\be
\delta V_{(1,1)}+\delta V_{(1,2)}=30 \frac{ a^2 A^2 e^{-2 a \tau_3}  \sqrt{\tau_3}}{c_1 \vo  \tau_1}-30 \frac{c_2  a A W_0 e^{-a \tau_3} \sqrt{\tau_3}}{c_1 \vo ^2 \sqrt{\tau_1}}- 90 \frac{a A e^{-a \tau_3}W_0 \tau_3}{\vo ^2 \tau_1}.
\ee
Our aim is to evaluate the size of these contributions the potential in the vicinity of the LVS minimum. While Eq. (\ref{eq:VLVSfibre}) determines that at the putative minimum $\tau_3\sim \log \vo$, the VEV for the fibre modulus $\tau_1$ depends on the chosen, and at this point unspecified, mechanism for its stabilisation. A universal bound on $\tau_1$ comes from the requirement of consistency of the supergravity approximation which leads to the constraint  $\tau_1> l_s^2$. Depending on the mechanism employed to lift the fibre modulus, one may identify two extreme limits: isotropic compactifications where $1<\tau_1<\tau_2$ and anisotropic compactifications where $1<\tau_1\ll\tau_2$. Approximating $\vo\sim \tau_2 \sqrt{\tau_1}/2$ we may rewrite these limits as:
\be
\text{isotropic:\: }1<\tau_1^{3/2}<3 \vo \qquad,\qquad \text{anisotropic:\: }1<\tau_1^{3/2}\ll3 \vo.
\label{eq:condFibre1}
\ee
With these hierarchies in mind we recast the $\alpha'^2$ corrections to the potential into the following form

\be
\delta V_{(1,0)}+\delta V_{(2,0)}\sim \frac{2025}{8} \frac{W_0^2}{\vo ^2 \tau_1^2}\left(1-  \frac{8}{3}\frac{  \tau_1^{3/2} }{ \vo}+  \frac{c_2^2}{3 c_1 \sqrt{a}} \frac{\tau_1 \sqrt{\log\vo}}{\vo}+ \frac{c_2}{a} \frac{ \log \vo \sqrt{\tau_1}}{\vo}+ \frac{c_1}{a^{3/2}}\frac{ (\log \vo )^{3/2}}{\vo}\right)\,,
\label{eq:VpureFibre}
\ee
\be
\delta V_{(1,1)}+\delta V_{(1,2)}\sim \frac{45}{2} \frac{c_2  W_0^2 \log \vo}{a \vo^3 \sqrt{\tau_1}}\left(1+\frac{15 c_1 \sqrt{\log \vo}}{4 c_2 \sqrt{a} \sqrt{\tau_1}}\right),
\ee
where we have used 
\be
e^{-a\langle\tau_3\rangle}= \frac{3 c_1}{4 }\frac{W_0 \sqrt{\langle\tau_3\rangle}}{a A \langle\vo\rangle}\left(1-\frac{3}{4 a \langle\tau_3\rangle}\right)\,,
\ee
at the vacuum, as follows from Eq. (\ref{eq:VLVSfibre}).

Regardless of the regime where one stabilises the fibre modulus, it is clear that the most dangerous correction to the LVS potential is the first term in Eq. (\ref{eq:VpureFibre}), being the less volume suppressed than $V_{LVS}$ and having an overall prefactor of $\mc{O}(10^2)$. We note that a similar term, albeit with an $\mc{O}(1)$ prefactor and with extra $g_s$ suppression, was also present in the analysis of \cite{Cicoli:2007xp}. In the context of \cite{Cicoli:2007xp} it was possible to get rid of such term by suitably choosing the brane configuration, here however, once we choose the geometry of the compact space to be that of  $\mathbb{CP}_{11226}^4[12]$, this term is inevitable. For the existence of the LVS minimum one must require the GSW term to be subleading when compared to
\be
\langle V_{LVS} \rangle \sim  -\frac{3 c_1}{4} \frac{W_0^2 }{a^{3/2}}\frac{\sqrt{\log\langle \vo\rangle}}{\langle\vo\rangle^3}.
\ee
As in the case of $\mathbb{CP}_{11169}^4[18]$ one defines the ratio
\be
\Delta\equiv\frac{\delta V}{V_{LVS}}\sim \frac{675 a^{3/2} \langle\vo\rangle }{2 c_1 \tau_1^2\sqrt{\log\langle\vo\rangle}}\,.
\label{eq:condFibre2}
\ee
By demanding equality between $\tau_1=(3\vo)^{2/3}$ as the upper limit from Eq~\eqref{eq:condFibre1} and
\begin{equation}
 \tau_1=\frac{15 \sqrt{\frac{3}{2}} a^{3/4} \sqrt{\langle\vo\rangle}}{\sqrt{c_1} \sqrt{\Delta } \sqrt[4]{\log \langle\vo\rangle}}\,,
\end{equation}
from Eq.~\eqref{eq:condFibre2}, we get an absolute lower bound on the volume necessary to control the correction
\be
\langle\vo\rangle\gtrsim 10^{9}\cdot \frac{\left(\frac{a}{2\pi }\right)^{9/2} c_1^{-3} \Delta^{-3}}{38+\frac{23}{5}\ln\left[\left(\frac{a}{2\pi }\right)^{3} c_1^{-2}\Delta^{-2}\right]}
\ee

In Table \ref{tab:volumeBounds2} we present the lower limit on the volume of the compact space for different values of $a $ and $\Delta$, setting $c_1=1$.
\begin{table}[h!]
\begin{center}
\begin{tabular}{l ||c|c}
&$a=2\pi$&$a=2\pi/10$\\
\hline
\hline
$\Delta=1$ & $\sim 10^{7} $ & $5\times 10^{3}  $ \\
\hline
$\Delta=10^{-1}$ & $\sim 10^{10}  $ & $\sim 10^{6}  $ \\
\hline
$\Delta=10^{-2}$ & $\sim 10^{13}  $ & $\sim 10^{9}  $ \\
\end{tabular}
\end{center}
\caption{Bounds on the volume for LVS on $\mathbb{CP}_{11226}^4[12]$. }
\label{tab:volumeBounds2}
\end{table}

There are two interesting facts to note about this setup: Firstly, isotropic compactifications are less problematic  than anisotropic ones as they are characterised by larger VEVs for $\tau_1$. Secondly,  these limits on $\vo$ are independent from the way the extra small modulus contributes to the GSW correction.

\begin{figure}[h]
\begin{center}
\includegraphics[width=0.45\textwidth]{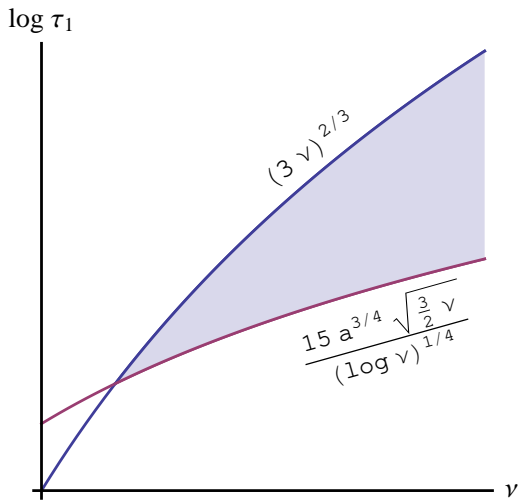}
\caption{Schematical validity of the LVS potential for the fibered manifold  $\mathbb{CP}_{11226}^4[12]$: the LVS minimum exists in the parts of the shaded are that are parametrically above the red line. The two curves intersect at around $\vo\sim 10^{7}$ for $\Delta=1$ and $a=2\pi$.}
\label{fig:FibrePlot}
\end{center}
\end{figure}

Just like in the $\mathbb{CP}_{11169}^4[18]$, the $\alpha'^2$ contributions to $V$ tend to dominate over the LVS potential, unless the volume is made sufficiently large. This in turn puts models that require smaller volumes, like modular inflation, under severe strain.  As one sees from Table~\ref{tab:volumeBounds2}, modular inflation models~\cite{Conlon:2005jm,Westphal:2005yz,Cicoli:2008gp} do not work in the most generic version of the Large Volume Scenario with just ED3-branes, but require the presence of gaugino condensation on stacks of D7-branes with gauge group ranks $\gtrsim {\cal O}(10)$.

This requirement becomes even more severe as we expect the $\alpha'^2$-correction to completely destroy the slow-roll properties of any inflationary model at $\Delta\lesssim {\cal O}(1)$ since such values of $\Delta$ are just sufficient to maintain a meta-stable minimum. Slow-roll most likely requires $\Delta\lesssim 0.01$ which renders volumes small enough for modular inflation impossible except if involving distastefully large-rank gaugino condensation with ranks $\gtrsim {\cal O}(\text{a few times}\;100)$.


\section{Conclusions}\label{sec:concl}
In this paper, we have analysed a new ${\cal N}=1$ string tree level correction at ${\cal O}(\alpha'^2)$ to the K\"ahler potential of the volume moduli of type IIB Calabi-Yau flux compactification found recently by Grimm, Savelli, and Weissenbacher~\cite{Grimm:2013gma}. We have looked at the structure of the GSW correction in the 4D volume moduli K\"ahler potential for general CY 3-folds with several K\"ahler moduli, in particular for CYs with a classical volume of approximate `swiss-cheese' form. As examples we took the scalar potential for a general 1-parameter KKLT  and K\"ahler uplifting scenarios, the 2-parameter `swiss-cheese' CY $\mathbb{CP}^4_{11169}$ in the Large Volume Scenario (LVS) and the K\"ahler uplifting scenario, and an anisotropic 3-parameter case in the context of LVS based on the fibered CY $\mathbb{CP}^4_{11226}$ relevant e.g. for the model of fibre inflation~\cite{Cicoli:2008gp}, respectively.

In all cases, we imposed as a stability criterion that the relative magnitude $\Delta\equiv \delta V / V_0$ of the correction in the scalar potential $\delta V$ compared to the scalar potential $V_0$ employed in the given stabilisation mechanism should be somewhat small $\Delta < 1$, typically $\Delta \lesssim 0.1$. This leads to an \emph{upper} bound $|W_0|\lesssim 10^{-3}$ on the flux superpotential in KKLT, and a \emph{lower} bound $\vo\gtrsim 10^8\ldots 10^9$ on the CY volume in LVS for the minima to persist in presence of the GSW correction for the most generic situation where the non-perturbative effect needed in LVS arises from Euclidean D3-brane instantons. The latter bound implies in particular, that the models of K\"ahler moduli inflation~\cite{Conlon:2005jm,Westphal:2005yz} including fibre inflation~\cite{Cicoli:2008gp} are under serious pressure from the new ${\cal O}(\alpha'^2)$-correction, since the bound $\vo > 10^8$ is difficult to reconcile with COBE normalization of the curvature 
perturbation and maintaining slow-roll flatness.


We find moreover, that the method of K\"ahler uplifting can operate in the presence of the new correction, if one manages to stabilise the volume at values $\vo \gtrsim 10^3$. This is a serious constraint for model building, as such volumes require either a distastefully large-rank condensing gauge group on a D7-brane stack~\cite{Rummel:2011cd}, or a double-gaugino condensate racetrack sector~\cite{Kallosh:2004yh,Sumitomo:2013vla}.

The above constraints should generalize to all type IIB CY flux compactifications, except for those where the correction depends just on a combination of the small blow-up volume moduli for which we give a criterion in Section~\ref{sec:GS}.

\section*{Acknowledgments}
We would like to thank T. W. Grimm, I. Ben-Dayan, J. P. Conlon, S. Kachru, A. Linde, R. Savelli, G. Shiu, E. Silverstein, M. Torabian, Y. Sumitomo, and H. Tye for interesting discussions and correspondence. MR would like to thank the Hong Kong University of Science and Technology where part of this work was completed for their warm hospitality. AW would like to thank KITP and the organizers and staff of the Primordial Cosmology 2013 workshop for their warm hospitality, congenial atmosphere, and support, where part of this work was completed. This research was supported in part by the Impuls und Vernetzungsfond of the Helmholtz Association of German Research Centres under grant HZ-NG-603, and by the National Science Foundation under Grant No. NSF PHY11-25915.

\appendix
\section{Corrections to no-scale \label{Eqs}}

The remaining terms in Eq. (\ref{eq:Vexpanded}) are
\be
\delta V_{(0,1)}\equiv 2\ e^K\big|_0 (W_0 \overline{\delta W}_a+ c.c.)  K^0_b K_0^{ab},
\label{eq:dV01}
\ee
\be
\begin{split}
\delta V_{(1,1)}\equiv &e^K\big|_0 \left \{K_0^{ab}\left[ 2 \ \delta K_b (\delta W_a \overline{W_0}+c.c.)+2 \ K^0_a \delta K_b (W_0\overline{\delta W}+c.c.)\right]\right.\\
& \left.-K_0^{am} \delta K_{mn} K_0^{n b}\left[ 2 \ (\delta W_a \overline{W_0}+c.c.)K^0_b +K^0_a K^0_b (W_0\overline{\delta W}+c.c.)\right]\right\}+  \frac{e^K\big|_1}{e^K\big|_0} \delta V_{(0,1)},
\end{split}
\label{eq:dV11}
\ee
\be
\begin{split}
\delta V_{(2,0)}\equiv & e^K\big|_0 |W_0|^2 \left\{\delta K_a K_0^{ab} \delta K_b-2\ K^0_a K_0^{am}\delta K_{mn} K_0^{nb} \delta K_b + K^0_a K_0^{am}\delta K_{mn} K_0^{np} \delta K_{p q} K_0^{q b }K^0_b \right\}+\\
 & -e^K\big|_1 |W_0|^2 \left\{-2\ K^0_a K_0^{ab} \delta K_b+K^0_a K_0^{am}\delta K_{mn} K_0^{nb} K^0_b\right\} + \frac{e^K\big|_2}{e^K\big|_0}V_{(0,0)}, 
\end{split}
\label{eq:dV20}
\ee
\be
\delta V_{(0,2)}\equiv  e^K\big|_0 K_0^{ab}\left[ 4\ \delta W_a\overline{\delta W}_b+2\ (\delta W_a \overline{\delta W}+c.c.) K^0_b\right],
\label{eq:dV02}
\ee
and finally
\be
\begin{split}
\delta V_{(1,2)}=&e^K\big|_0\left[ K_0^{ab} (2\ (\delta W_a \overline{\delta W}+c.c.)\delta K_b+2\ |\delta W|^2 K_a^0 \delta K_b)\right.\\
&\left.-K_0^{am}\delta K_{mn} K_0^{nb} \left(4\ \overline{\delta W}_a  \delta W_b+2\ (\delta W_a \overline{\delta W}+c.c)K^0_b+|\delta W|^2 K_a^0 K_b^0\right)\right]  + \frac{e^K\big|_1}{ e^K\big|_0}\delta V_{(0,2)}
\end{split}
\ee
Making repeated use of Eqs.  (\ref{eq:K0})  and (\ref{eq:noscale}) these can be written, without any further assumptions about the nature of $\delta K$ and $\delta W$, as 
\be
\delta V_{(0,1)}=-\frac{2 \tau_a}{\vo^2} (W_0 \overline{\delta W}_a+ c.c.) ,
\ee
\be
\begin{split}
\delta V_{(1,1)}=&\frac{1}{\vo^2}\left\{2 \ (\delta W_a \overline{W_0}+c.c.)(K_0^{ab}\delta K_b +K_0^{a m} \delta K_{mn}\tau_n)\right.\\
&\left.-(W_0\overline{\delta W} +c.c.)(2\ \tau_b \delta K_b+\tau_m \delta K_{m n}\tau_n)\right\}-2 e^K\big|_1 (W_0 \overline{\delta W}_a+ c.c.) \tau_a,
\end{split}
\ee
\be
\begin{split}
\delta V_{(2,0)}=&\frac{|W_0|^2}{\vo^2} K_0^{ab}\left\{\delta K_a \delta K_b +2 \ \tau_m \delta K_{m a} \delta K_b+\tau_m \delta K_{ma} \delta K_{b q} \tau_q\right\}\\&-e^K\big|_1 |W_0|^2 \left\{2\ \tau_b \delta K_b+\tau_m \delta K_{mn} \tau_n \right\}\,, 
\end{split}
\ee
\be
\delta V_{(0,2)}\equiv  \frac{1}{\vo^2}\left[4\ \delta W_a K_0^{ab}\overline{\delta W_b} -2\ \tau_a (\delta W_a \overline{\delta W}+c.c.)\right],
\ee
\be
\begin{split}
\delta V_{(1,2)}=&\frac{1}{\vo^2}\left[2\ (\delta W_a \overline{\delta W}+c.c.)(K_0^{ab}\delta K_b+K_0^{am}\delta K_{mn}\tau_n)-|\delta W|^2 (2\ \tau_b \delta K_b+\tau_m \delta K_{mn}\tau_n)\right.\\
&\left.-4\ \delta W_a\overline{\delta W_b} K_0^{am}\delta K_{mn}K_0^{bn}\right]+e^K\big|_1[4\ K_0^{ab}\delta W_a \overline{\delta W_b}-2 \tau_a(\delta W_a \overline{\delta W}+c.c.)]\,.
\end{split}
\ee

\section{Calculation of the quantum corrected K\"ahler potentials}\label{sec:app}

In this section, we calculate the corrected K\"ahler potentials of Eq.~\eqref{K11169} and Eq.~\eqref{K11226} as the K\"ahler potential in the one-modulus case Eq.~\eqref{Ksingle} was already calculated in~\cite{Grimm:2013gma}.

\subsection{$\mathbb{CP}_{11169}^4[18]$}

The CY threefold $X_3$ can be given as an hypersurface in the ambient space $X^{\text{amb}}_{4}$ defined by the weight matrix
\begin{equation}
X^{\text{amb}}_{4} :\quad
\begin{array}{cccccc}
u_1 & u_2 & u_3 & u_4 & u_5 & \xi\\ 
\hline 
1 & 1 & 1 & 6 & 0 & 9\\
0 & 0 & 0 & 2 & 1 & 3\\
\end{array}\quad \:.
\label{WSX411169}
\end{equation}
The defining equation for the CY is
\begin{equation}
 \xi^2 = P_{18,6}(u_1,..,u_5)\,,
\end{equation}
where $\xi \to -\xi$ defines the orientifold involution. The toric divisors $D_i$ are defined as hypersurfaces with complex codimension one via $u_i=0$. From the weights in Eq.~\eqref{WSX411169}, we can read of that $D_1=D_2=D_3$ and $D_4=6 D_1+2 D_5$. The base $B_3$ of the F-theory uplift can be written as
\begin{equation}
B_3 :\quad
\begin{array}{ccccc}
u_1 & u_2 & u_3 & u_4 & u_5\\ 
\hline 
1 & 1 & 1 & 6 & 0\\
0 & 0 & 0 & 2 & 1\\
\end{array}\quad \:.
\label{WSB311169}
\end{equation}
We pick the two forms $\hat D_4$ and $\hat D_5$ as a basis of $H^{1,1}(X_3,\mathbb{Z})$, hence $J=t_4 D_4 + t_5 D_5$. The first Chern class of the base is given as
\begin{equation}
 c_1(B_3) = 3 D_1 + D_4 + D_5 = \frac{3}{2} D_4\,,
\end{equation}
using $D_1 = \frac{1}{6}D_4-\frac{1}{3}D_5$. The triple intersections of $X_3$ can be computed using \textit{PALP}~\cite{Kreuzer:2002uu,Braun:2012vh} to be $\kappa_{444}=72$ and $\kappa_{555}=9$ while all other intersection numbers vanish. Hence, the classical volume is given as
\begin{equation}
 \vo = 12 t_4^3 + \frac32 t_5^3\,,
\end{equation}
The K\"ahler cone is given by
\begin{equation}
 2 t_4 > -t_5 \qquad \text{and} \qquad -t_5 >0\,,
\end{equation}
such that inverting the relation between 4- and 2-cycle volumes, Eq.~\eqref{eq:24cycles} yields
\begin{equation}
 t_4 =  \frac16 \sqrt{\tau_4} \qquad \text{and} \qquad t_5 = -\frac{\sqrt{2}}{3}  \sqrt{\tau_5}\,,
\end{equation}
such that
\begin{equation}
 \vo = \frac{1}{18} \tau_4^{3/2}-\frac{\sqrt{2}}{9} \tau_5^{3/2}\,.\label{Vol11169app}
\end{equation}
The $\alpha'^2$ correction term is given by
\begin{equation}
 \int_{X_3} c_1(B_3)^2 \wedge J = 162 t_4 = 27 \sqrt{\tau_4}\,,\label{corr11169app}
\end{equation}
using the triple intersection numbers of $X_3$. Combining Eq.~\eqref{Vol11169app} and Eq.~\eqref{corr11169app} gives the corrected K\"ahler potential of Eq.~\eqref{K11169} for $\tau_1 \equiv \tau_4$ and $\tau_2 \equiv \tau_5$.

Note that there is a peculiar relation between the volumes of the base $B_3$ and the CY $X_3$ in the case of $\mathbb{CP}_{11169}^4[18]$ that was already discussed extensively in~\cite{Louis:2012nb}: The triple self intersection $D_5 \cap D_5 \cap D_5$ lies on the 4-cycle defined by $\xi = 0$, i.e., the cycle that is wrapped by the O7 plane. Hence, the intersection in the double cover $X_3$ is not twice the intersection in $B_3$ as is the case for an intersection away from the O7 plane. The base volume can be determined to be
\begin{equation}
 \vo_{B_3} = \frac{\sqrt{2}}{18} \left( \tau_4^{3/2}- \tau_5^{3/2} \right)\,,
\end{equation}
which is not simply $\vo / 2$ as given in Eq.~\eqref{Vol11169app}.

\subsection{$\mathbb{CP}_{11226}^4[12]$}

In this example, the CY threefold $X_3$ can be given as an hypersurface 
\begin{equation}
 \xi^2 = P_{12,6}(u_1,..,u_5)\,,
\end{equation}
in the ambient space $X^{\text{amb}}_{4}$ defined by the weight matrix
\begin{equation}
X^{\text{amb}}_{4} :\quad
\begin{array}{cccccc}
u_1 & u_2 & u_3 & u_4 & u_5 & \xi\\ 
\hline 
1 & 1 & 2 & 2 & 0 & 6\\
0 & 0 & 1 & 1 & 1 & 3\\
\end{array}\quad \:.
\label{WSX411226}
\end{equation}
From the weights in Eq.~\eqref{WSX411169}, we can read of that $D_1=D_2$ and $D_3=D_4=2 D_1+ D_5$. The base $B_3$ of the F-theory uplift can be written as
\begin{equation}
B_3 :\quad
\begin{array}{ccccc}
u_1 & u_2 & u_3 & u_4 & u_5\\ 
\hline 
1 & 1 & 2 & 2 & 0\\
0 & 0 & 1 & 1 & 1\\
\end{array}\quad \:.
\label{WSB311226}
\end{equation}
Pick the two forms $\hat D_1$ and $\hat D_4$ as a basis of $H^{1,1}(X_3,\mathbb{Z})$, we find
\begin{equation}
 c_1(B_3) = 2 D_1 + 2 D_4 + D_5 = 3 D_4\,,
\end{equation}
using $D_5 = D_4 - 2 D_1$. The only non-vanishing triple intersection numbers are $\kappa_{144}=2$ and $\kappa_{444}=4$, such that the relation between the relation between the 2-cycles and 4-cycles is
\begin{equation}
 t_1 = \frac{\tau_4-2 \tau_1}{2 \sqrt{\tau_1}} \qquad \text{and} \qquad t_4 = \sqrt{\tau_1}\,.
\end{equation}
Then, we can calculate
\begin{equation}
 \vo = \frac16 \sqrt{\tau_1} (3\tau_4 -2 \tau_1)\,,\label{Vol11226app}
\end{equation}
and
\begin{equation}
 \int_{X_3} c_1(B_3)^2 \wedge J = 18 t_1 + 36 t_4 = 9\frac{2 \tau_1 + \tau_4}{\sqrt{\tau_1}}\,.\label{corr11226app}
\end{equation}
Up to the blow-up modulus, Eq.~\eqref{Vol11226app} and Eq.~\eqref{corr11226app} determine the quantum corrected K\"ahler potential in Eq.~\eqref{K11226} for $\tau_2 \equiv \tau_4$.


\begin{thebibliography}{10}

\bibitem{Grimm:2013gma} 
  T.~W.~Grimm, R.~Savelli and M.~Weissenbacher,
  arXiv:1303.3317 [hep-th].

\bibitem{Douglas:2006es} 
  M.~R.~Douglas and S.~Kachru,
  Rev.\ Mod.\ Phys.\  {\bf 79}, 733 (2007)
  [hep-th/0610102].



\bibitem{Grana:2005jc} 
  M.~Grana,
  Phys.\ Rept.\  {\bf 423}, 91 (2006)
  [hep-th/0509003].



\bibitem{Blumenhagen:2006ci} 
  R.~Blumenhagen, B.~Kors, D.~Lust and S.~Stieberger,
  Phys.\ Rept.\  {\bf 445}, 1 (2007)
  [hep-th/0610327].


\bibitem{Kachru:2003aw} 
  S.~Kachru, R.~Kallosh, A.~D.~Linde and S.~P.~Trivedi,
  Phys.\ Rev.\ D {\bf 68}, 046005 (2003)
  [hep-th/0301240].



\bibitem{Susskind:2003kw} 
  L.~Susskind,
  In *Carr, Bernard (ed.): Universe or multiverse?* 247-266
  [hep-th/0302219].



\bibitem{Giddings:2001yu} 
  S.~B.~Giddings, S.~Kachru and J.~Polchinski,
  Phys.\ Rev.\ D {\bf 66}, 106006 (2002)
  [hep-th/0105097].



\bibitem{Dasgupta:1999ss} 
  K.~Dasgupta, G.~Rajesh and S.~Sethi,
  JHEP {\bf 9908}, 023 (1999)
  [hep-th/9908088].



\bibitem{Saltman:2004jh} 
  A.~Saltman and E.~Silverstein,
  JHEP {\bf 0601}, 139 (2006)
  [hep-th/0411271].



\bibitem{Parameswaran:2006jh} 
  S.~L.~Parameswaran and A.~Westphal,
  JHEP {\bf 0610}, 079 (2006)
  [hep-th/0602253].



\bibitem{Silverstein:2007ac} 
  E.~Silverstein,
  Phys.\ Rev.\ D {\bf 77}, 106006 (2008)
  [arXiv:0712.1196 [hep-th]].



\bibitem{Burgess:2003ic} 
  C.~P.~Burgess, R.~Kallosh and F.~Quevedo,
  JHEP {\bf 0310}, 056 (2003)
  [hep-th/0309187].



\bibitem{Lebedev:2006qq} 
  O.~Lebedev, H.~P.~Nilles and M.~Ratz,
  Phys.\ Lett.\ B {\bf 636}, 126 (2006)
  [hep-th/0603047].



\bibitem{Balasubramanian:2005zx} 
  V.~Balasubramanian, P.~Berglund, J.~P.~Conlon and F.~Quevedo,
  JHEP {\bf 0503}, 007 (2005)
  [hep-th/0502058].



\bibitem{Becker:2002nn} 
  K.~Becker, M.~Becker, M.~Haack and J.~Louis,
  JHEP {\bf 0206}, 060 (2002)
  [hep-th/0204254].



\bibitem{Balasubramanian:2004uy} 
  V.~Balasubramanian and P.~Berglund,
  JHEP {\bf 0411}, 085 (2004)
  [hep-th/0408054].



\bibitem{Rummel:2011cd} 
  M.~Rummel and A.~Westphal,
  JHEP {\bf 1201}, 020 (2012)
  [arXiv:1107.2115 [hep-th]].



\bibitem{Louis:2012nb} 
  J.~Louis, M.~Rummel, R.~Valandro and A.~Westphal,
  JHEP {\bf 1210}, 163 (2012)
  [arXiv:1208.3208 [hep-th]].



\bibitem{Berg:2005ja} 
  M.~Berg, M.~Haack and B.~Kors,
  JHEP {\bf 0511}, 030 (2005)
  [hep-th/0508043].



\bibitem{Berg:2007wt} 
  M.~Berg, M.~Haack and E.~Pajer,
  JHEP {\bf 0709}, 031 (2007)
  [arXiv:0704.0737 [hep-th]].



\bibitem{Cicoli:2007xp} 
  M.~Cicoli, J.~P.~Conlon and F.~Quevedo,
  JHEP {\bf 0801}, 052 (2008)
  [arXiv:0708.1873 [hep-th]].



\bibitem{Cicoli:2008gp} 
  M.~Cicoli, C.~P.~Burgess and F.~Quevedo,
  JCAP {\bf 0903}, 013 (2009)
  [arXiv:0808.0691 [hep-th]].



\bibitem{Conlon:2005jm} 
  J.~P.~Conlon and F.~Quevedo,
  JHEP {\bf 0601}, 146 (2006)
  [hep-th/0509012].



\bibitem{Westphal:2005yz} 
  A.~Westphal,
  JCAP {\bf 0511}, 003 (2005)
  [hep-th/0507079].



\bibitem{Kallosh:2004yh} 
  R.~Kallosh and A.~D.~Linde,
  JHEP {\bf 0412}, 004 (2004)
  [hep-th/0411011].


\bibitem{Sumitomo:2013vla} 
  Y.~Sumitomo, S.~-H.~H.~Tye and S.~S.~C.~Wong,
  arXiv:1305.0753 [hep-th].
  
\bibitem{Cremmer:1983bf} 
  E.~Cremmer, S.~Ferrara, C.~Kounnas and D.~V.~Nanopoulos,
  Phys.\ Lett.\ B {\bf 133}, 61 (1983).



\bibitem{Ellis:1983sf} 
  J.~R.~Ellis, A.~B.~Lahanas, D.~V.~Nanopoulos and K.~Tamvakis,
  Phys.\ Lett.\ B {\bf 134}, 429 (1984).



\bibitem{Sen:1997gv} 
  A.~Sen,
  Phys.\ Rev.\ D {\bf 55}, 7345 (1997)
  [hep-th/9702165].
  
\bibitem{Collinucci:2008pf} 
  A.~Collinucci, F.~Denef and M.~Esole,
  JHEP {\bf 0902}, 005 (2009)
  [arXiv:0805.1573 [hep-th]].

\bibitem{Braun:2008ua} 
  A.~P.~Braun, A.~Hebecker and H.~Triendl,
  Nucl.\ Phys.\ B {\bf 800}, 298 (2008)
  [arXiv:0801.2163 [hep-th]].

\bibitem{Kreuzer:2000xy} 
  M.~Kreuzer and H.~Skarke,
  Adv.\ Theor.\ Math.\ Phys.\  {\bf 4}, 1209 (2002)
  [hep-th/0002240].



\bibitem{Gray:2012jy} 
  J.~Gray, Y.~-H.~He, V.~Jejjala, B.~Jurke, B.~D.~Nelson and J.~Simon,
  Phys.\ Rev.\ D {\bf 86}, 101901 (2012)
  [arXiv:1207.5801 [hep-th]].

\bibitem{MayrhoferThesis}
C.~Mayrhofer, {\it {"Compactifications of Type IIB String Theory and F-Theory
  Models by Means of Toric Geometry"}},  {\em PhD thesis, Vienna University of
  Technology, 11} (2010).

\bibitem{Westphal:2006tn} 
  A.~Westphal,
  JHEP {\bf 0703}, 102 (2007)
  [hep-th/0611332].



\bibitem{Bond:2006nc} 
  J.~R.~Bond, L.~Kofman, S.~Prokushkin and P.~M.~Vaudrevange,
  Phys.\ Rev.\ D {\bf 75}, 123511 (2007)
  [hep-th/0612197].



\bibitem{Cicoli:2011ct} 
  M.~Cicoli, F.~G.~Pedro and G.~Tasinato,
  JCAP {\bf 1112}, 022 (2011)
  [arXiv:1110.6182 [hep-th]].



\bibitem{Cicoli:2011yy} 
  M.~Cicoli, C.~P.~Burgess and F.~Quevedo,
  JHEP {\bf 1110}, 119 (2011)
  [arXiv:1105.2107 [hep-th]].
  
\bibitem{Cicoli:2012tz}
  M.~Cicoli, F.~G.~Pedro and G.~Tasinato,
  JCAP {\bf 1207} (2012) 044
  [arXiv:1203.6655 [hep-th]].



\bibitem{Kreuzer:2002uu} 
  M.~Kreuzer and H.~Skarke,
  Comput.\ Phys.\ Commun.\  {\bf 157}, 87 (2004)
  [math/0204356 [math-sc]].



\bibitem{Braun:2012vh} 
  A.~P.~Braun, J.~Knapp, E.~Scheidegger, H.~Skarke and N.~-O.~Walliser,
  arXiv:1205.4147 [math.AG].


\end{thebibliography}
\end{document}